\documentstyle[11pt]{article}

\makeatletter%
\def\nottoobig#1{{\hbox{$\left#1\vcenter to1.111\ht\strutbox{}\right.\n@space$}}}
\makeatother%

\makeatletter %

\newcommand{\implies}{\:\Rightarrow\:}

\newlength{\filength}
\newsavebox{\gcbox}
\sbox{\gcbox}{\framebox[\filength]{\rule{0ex}{2ex}}}

\newcommand{\red}[3]{ {  {\rm R}_{#1}^{#2}({#3}) }    }

\newcommand{\rms}{ {  {\rm R}_m^{\rm \#P}}}
\newcommand{\npclosure} {  \red{m}{\#{\rm P}}{\rm NP}}
\newcommand{\pclosure} {  \red{m}{\#{\rm P}}{\rm P}}

\newtheorem{theorem}{Theorem}[section]
\newtheorem{corollary}[theorem]{Corollary}
\newcommand{\qedblob}{\mbox{\rule[-1.5pt]{5pt}{10.5pt}}}
\def\literalqed{{\ \nolinebreak\hfill\mbox{\qedblob\quad}}}

\def\qed{\literalqed}

\newtheorem{proposition}{Proposition}
\makeatother%

\newcommand{\sharpp}{{\rm \#P}}
\newcommand{\sharpnp}{{\rm \#NP}}
\newcommand{\sharpph}{{\rm \#PH}}
\newcommand{\sharpsat}{{\rm \#SAT}}
\newcommand{\sat}{{\rm SAT}}

\newcommand{\parityp}{{\rm \oplus P}}
\newcommand{\up}{{\rm UP}}

\newcommand{\p}{{\rm P}}

\newcommand{\NP}{{\rm NP}}
\newcommand{\np}{{\rm NP}}

\newcommand{\numaccm}{  \#acc_M}

\newcommand{\pcalonetwo}{  {\rm P}^{ {\cal C}_1 [1] : {\cal C}_2 [1]}}
\newcommand{\pcalonetwoplus}{  {\rm P}^{ {\cal C}_1 [1] : {\cal C}_2 [1]_+}}

\newcommand{\psnplus}{  {\rm P}^{ {\rm \#P} [1] : {\rm NP} [1]_+}}

\newcommand{\psnnoplusbigohone}{  {\rm P}^{ {\rm \#P} [1] : {\rm NP} [{\cal O}(1)]}}

\newcommand{\pns}{  {\rm P}^{ {\rm NP} [1] : {\rm \#P} [1]}}
\newcommand{\psone}{  {\rm P}^{ {\rm \#P} [1]}}

\newcommand{\pp}{{\rm PP}}

\newcommand{\selfoutput}{{\rm SelfOutput}}
\newcommand{\selfpath}{{\rm SelfPath}}
\newcommand{\ceqp}{{\rm C_{\!=}P}}
\newcommand{\pspace}{{\rm PSPACE}}

\newcommand{\pnp}{{\p^{\rm NP}}}
\newcommand{\psharpp}{{\p^{\rm \#P}}}

\newcommand{\ph}{{\rm PH}}

\newcommand{\manyone}{\mbox{$\leq_m^p$}}

\newcommand{\sigmastar}{\mbox{$\Sigma^\ast$}}

\newcommand{\calf}{\mbox{$\cal F$}}
\newcommand{\calc}{\mbox{$\cal C$}}

\newcommand{\calone}{\mbox{${\cal C}_1$}}
\newcommand{\caltwo}{\mbox{${\cal C}_2$}}

\newcommand{\condition}{\,\nottoobig{|}\:}
\def\land{{\; \wedge \;}}

\newcommand{\manyonesharp}{\mbox{$\leq_m^{\sharpp}$}}

\newcommand{\redmsharp}[1]{ {\rm R}_m^{\sharpp}({#1})}

\newcommand{\naturalnumbers}{{\bf N}}
\begin{document}

\title{Self-Specifying Machines}

\author{
Lane A. Hemaspaandra\footnote{Supported in part 
by grants NSF-CCR-9322513 and 
NSF-INT-9513368/DAAD-315-PRO-fo-ab.  
Work done in part while visiting
Friedrich-Schiller-Universit\"at Jena.
Email: {\tt lane@cs.rochester.edu}.}
\\Department of Computer Science\\University of Rochester\\
            Rochester, NY 14627, USA
\and
Harald Hempel and Gerd Wechsung\footnote{Supported in part 
by grant
NSF-INT-9513368/DAAD-315-PRO-fo-ab.
Work done in part while 
visiting 
Le Moyne College.
Email: {\tt \{hempel,wechsung\}@\protect\linebreak[0]informatik.\protect\linebreak[0]uni-jena.de}.}
\\Institut f\"ur Informatik\\
Friedrich-Schiller-Universit\"at Jena\\
07740 Jena, Germany}

\date{October 22, 1998}

\maketitle

\begin{abstract}
We study the computational power of machines that specify their own 
acceptance types, and show that they accept exactly the languages 
that $\manyonesharp$-reduce to NP sets.
A natural variant accepts exactly the languages
that $\manyonesharp$-reduce to P sets.
We show that these two classes coincide if and only if 
$\psone = \psnnoplusbigohone$, 
where the latter class 
denotes the sets acceptable via at most one question to $\sharpp$ 
followed by at most 
a constant 
number of questions to $\np$.
\end{abstract}

\section{Introduction}

This paper studies the power of self-specifying 
acceptance types.
As is standard, by acceptance type we mean 
the set of numbers of accepting 
paths that cause a machine to accept.  Many complexity classes have a 
fixed acceptance type.  For example, a set is in NP if and only if 
for some nondeterministic polynomial-time Turing machine (NPTM) $M$
it holds that for each $x$, $x\in L$ if and only if $\numaccm(x) \in 
\{1,2,3,\cdots\}$, where $\numaccm(x)$ represents the number 
of accepting paths of machine $M$ on input
$x$.  Replacing $\{1,2,3,\cdots\}$ with the 
set $\{1,\,3,\,5,\,\cdots\}$ yields a perfectly acceptable 
definition of the complexity class 
$\parityp$~\cite{pap-zac:c:two-remarks,gol-par:j:ep},
and so on for many standard classes, such 
as coNP, US~\cite{bla-gur:j:unique-sat}, etc.  
In fact, quite surprisingly, it turns out that there is 
a {\em single\/} polynomial-time computable set,
${\rm MiddleBit} = \{ i \condition$ the 
$\lfloor (\log_2{i})/2\rfloor$th
bit of $i$ is a one$\}$, that is universal for $\rm \pp^{PH}$ 
in the sense that this one set serves simultaneously as the fixed
accepting type of all sets in $\rm \pp^{PH}$.  This follows 
immediately from the fact that 
$\rm \pp^{PH} \subseteq MP$~(Green 
et al.~\cite{gre-koe-reg-sch-tor:j:middle-bit},
who also define the ``middle bit'' class MP).
Thus, in standard notation (a full definition is included later), 
the fixed set 
MiddleBit has the property that $ {\rm \pp^{PH}} \subseteq 
{\rm R}_m^{\rm \#P}(\{{\rm MiddleBit}\})$.

In contrast, some classes have been defined via an {\em external\/} 
function or set specifying their acceptance type.  For example, the complexity 
class $\ceqp$~\cite{sim:thesis:complexity,wag:j:counting-recursion}
can be defined as the class of languages $L$ such that for some NPTM $M$ and
some FP function $g$, for all $x$, $x\in L$ if
and only if $\numaccm(x) = g(x)$.
In the even more abstract setting of so-called leaf 
languages~(\cite{bov-cre-sil:j:uniform,ver:j:oracle-survey},
see 
also~\cite{her-lau-sch-vol-wag:c:bit-reductions,jen-mck-the:j:leaf}),
separate predicates specify which numbers of accepting paths of a machine 
specify acceptance and which specify rejection.  Such notions are 
sufficiently flexible to describe a broad range of classes,
including even ``promise'' counting classes such 
as SPP~\cite{hem-ogi:j:closure,fen-for-kur:j:gap}, etc.

In this paper, we 
introduce and study self-specifying acceptance types.  A language is in 
SelfOutput if for 
some NPTM $M$ and every input $x$, $x\in L$ if
and only if $
\numaccm(x)$ is the output of some accepting 
path of $M(x)$.  Similarly, a language is in 
SelfPath if for 
some NPTM $M$ whose computation tree is always
a complete tree (of polynomial depth specified by
the input's length) and every input $x$, $x\in L$ 
if and only if 
the lexicographically $(\numaccm(x))$th path
of $M(x)$ is an accepting path.  Note that self-specification allows
the machine to dynamically set its own acceptance type, 
but restricts the machine
by requiring that the machine's paths not only specify
its acceptance type but also execute it (in the sense 
that they themselves form the $\numaccm$ set---i.e., the \#P function
in the sense defined below---being
analyzed by the acceptance type).

Valiant~\cite{val:j:permanent,val:j:enumeration}
introduced the class \#P, which is the class of functions $f$
such that for some NPTM $M$ and every $x$, $f(x)=\numaccm(x)$.
We prove that $\selfoutput = \npclosure$ and that $\selfpath = 
\pclosure$.  That is, SelfOutput and SelfPath 
consist of
the sets
that 
$\leq_m^{\#\rm P}$-reduce to NP and P sets, respectively.
Put another way, SelfOutput and SelfPath 
capture the power 
of counting 
with respect to 
NP-computable and P-computable
acceptance sets.
Essentially equivalently, we establish that $\selfoutput = \np//\sharpp$
and 
$\selfpath = \p//\sharpp$, where the $//$ is a certain recently introduced 
advice notation. 
For a definition of $//$ see Section~\ref{section:prelim}.
Also equivalently, we note
that $\selfoutput = \psnplus$, the class of languages accepted by
P machines given at most one call to a \#P oracle followed 
by at most one positive~\cite{lad-lyn-sel:j:com,sel:j:reductions-pselective}
query to an NP oracle.

Note that it is not at all clear whether 
$\selfpath$ equals $\selfoutput$ (equivalently, 
in light of the characterizations of 
this paper, whether $\psone = \psnplus$).  
However, we show
that these two classes are equal if and only if $\psone =
\psnnoplusbigohone$.  This is a so-called ``downward separation''
result (see, e.g., 
\cite{hem-jha:j:defying}, for some background),
and indeed what our proof actually establishes is
that the following three conditions are equivalent: 
\begin{enumerate}
\item $\psone =
\psnplus$,
\item $\psone = \p^{\sharpp[1]:\np[1]}$, and
\item $\psone =
\psnnoplusbigohone$.
\end{enumerate}

Since, in contrast with the just-mentioned open issue
of whether $\psone = \psnplus$, it is easy to see via  standard 
techniques~\cite{cai-hem:j:parity,pap-zac:c:two-remarks}
that 
$\psone$
{\em does\/} equal 
$\pns$ 
(indeed, even 
$\psone = \p^{ {\rm NP} [{\cal O}(\log n)]: {\rm \#P}[1]})$,
the comments of the previous paragraph give some weak 
evidence that 
order of access may be important in determining computational 
power, a theme that has been raised and studied 
in other settings (see the 
survey~\cite{hem-hem-hem:j:query-order-survey}).
Unfortunately, in the present setting, giving 
firm evidence for this seems hard.  In fact, there is no known
oracle separation of 
$\psone$ from PSPACE (much less 
of $\psone$
from $\psnplus$), though much effort has been made in 
that direction.

\section{Preliminaries}\label{section:prelim}

For standard notations and definitions that are not 
included here, we refer the reader to any complexity textbook,
e.g.,~\cite{bov-cre:b:complexity,bal-dia-gab:b:sctI-2nd-ed,pap:b:complexity}.
$\naturalnumbers = \{0,1,2,\cdots\}$.
For any $n \in \naturalnumbers$, define $string(n)$ to be $n$
written in binary with no leading zeros.  For any string $x\in \sigmastar$,
let $string(x) = x$.
For any $x \in \sigmastar - \epsilon$, define $int(x)$ to be $x$ 
interpreted as the binary representation of a natural number.
By convention, let $int(\epsilon) = 0$.  
(The function $int(\cdot)$ is not a one-to-one function,
e.g., $int(0011) = int(11) = 3$.  
It merely is the natural direct reading of strings as 
representations of natural numbers.)
For each NPTM $M$, for each $x\in \sigmastar$, 
and for each accepting path $y$ of $M(x)$, let
$pathout_M(x,y)$ be the integer output on path $y$, which by definition we 
take to be $int(w)$, 
where $w$ is the bits of the work tape between the left endmarker 
and
the first tape cell that holds neither a~0 nor a~1, i.e., 
that is some other symbol or a blank 
(see~\cite{boo-lon-sel:j:qual,boo-lon-sel:j:quant}).
Define $acc_M(x)=\{y  \in \sigmastar
\condition y$ is an accepting path of $M(x)\}$, 
$iacc_M(x) = \{ int(z) \condition z \in acc_M(x)\}$,
$iout_M(x)=\{ n \in \naturalnumbers \condition (\exists y) 
[y \in acc_M(x)$  and $n=pathout_M(x,y)]\}$, 
$\numaccm(x)=||acc_M(x)||$, and $span_M(x)=||iout_M(x)||$.
Recall that 
$\#{\rm P} = \{ f:\sigmastar \rightarrow \naturalnumbers \condition 
(\exists$ NPTM $M)(\forall x)[f(x) = 
\numaccm(x)]\}$~\cite{val:j:permanent,val:j:enumeration}.
$\# \sat$ is the function such that $\#\sat(f)$ is the number
of satisfying assignments of $f$ if $f$ is a satisfiable 
boolean formula, and $\#\sat(f)$ is~0 otherwise.
$\#\sat$ is known to be 
$\#{\rm P}$-complete
(see~\cite{val:j:permanent,val:j:enumeration,zan:j:sharp-p}).

A set $A$ is many-one reducible to $B$ via a $\sharpp$ 
function, $A \leq_m^{\sharpp} B$, if and only if there exists 
a function $f \in \sharpp$ such that, for all $x$, 
$x \in A \iff string(f(x)) \in B$.
For any $a$ and $b$ for which $\leq_a^b$
is defined and any class $\cal C$, let $ {\rm R}_a^b({\cal C}) =
\{L\condition (\exists C \in {\cal C})[L\leq_a^b C]\}$.

Hemaspaandra, Hempel, and Wechsung~\cite{hem-hem-wec:j:query-order-bh}
introduced the study of the power of ordered query access
(other papers on or related to query order 
include
\cite{hem-hem-hem:j:query-order-ph,agr-bei-thi:c:modulo-information,her:c:transformation-monoid-acceptance,bei-cha:c:commutative-queries,wag:j:parallel-difference,mcn:unpub:commutative-oracles,hem-hem-hem:j:sn-1tt-np-completeness} 
and the 
survey~\cite{hem-hem-hem:j:query-order-survey}).
We adopt this notion, 
and extend it to the 
function class case and to the positive query case.
For any function or language classes $\calone$ and $\caltwo$, define
$\pcalonetwo$ to be the class of languages accepted by polynomial-time
machines making at most 
one query to a $\calone$ oracle followed by at most one query to
a $\caltwo$ oracle.
If $\caltwo$ is a language class, then 
$\pcalonetwoplus$ is the class of all sets in $\pcalonetwo$ witnessed by 
a polynomial-time oracle machine that accepts if and only if 
the $\caltwo$ set is queried and the answer to 
that query is 
``yes.''\footnote{\protect\label{f:lg}{}We mention
that this requires that 
the querying machine is such that if (in light of
of the $\calone$ query's actual answer if any---note that we do 
not require that a query to either $\calone$ or $\caltwo$ 
necessarily be made) there is 
a $\caltwo$ query then
the 
truth-table the querying machine has with respect to the 
answer from the $\caltwo$ query, {\em given the true answer to 
the $\calone$ query if there is any such query\/}, is that the machine 
will accept if and only if the answer is ``yes.''  However, for
most natural classes, in particular $\caltwo = \np$, without
loss of generality we can assume that the querying machine always 
makes exactly two queries and that its truth table with respect
to the two answers (even given a ``lying'' answer to the first query)
is: accept if and only if the second answer is ``yes.''}

Two other formalisms can represent similar notions.  
Let $\langle \cdot , \cdot \rangle$ be a pairing function from
$\sigmastar \times \sigmastar$ to $\sigmastar$ having the standard
properties (easily computable, easily invertible, etc.).  
Recall that, as
noted above for the case ${\cal F} = \sharpp$, for any
class of functions $\calf$ and any sets $A$ and $B$:
$A \leq_m^{\cal F} B \iff (\exists  f\in \calf)(\forall x)[x\in A \iff
string(f(x)) \in B]$.  Generalizing the seminal ``advice classes'' notion
of Karp and Lipton~\cite{kar-lip:c:nonuniform}, K\"obler 
and Thierauf~\cite{koe-thi:j:opt} have 
studied an interesting notion, which previously appeared 
in less general form in work of Krentel and others.
K\"obler 
and Thierauf note that the notion is related 
to many-one reductions via functions.
The notion is as follows:
For any function class
$\cal F$ and any complexity class $\cal C$ define:
${\cal C}//\calf = \{L\condition (\exists f \in \calf)(\exists C\in 
\calc)(\forall x)[x\in L \iff \langle x,\,string(f(x))\rangle \in 
C ]\}$~\cite{koe-thi:j:opt}.

We note that it will not always be the case that $\p^{ {\cal F}[1]:
{\cal C}[1]_+} = \calc // \calf = {\rm R}_m^{\cal F}(\calc)$.  The
reason why this may not always hold 
is that it is possible that $\calc$ lacks the power to
decode pairing functions or to form from a function value
the appropriate $\calc$ query, or that (in terms of proving $\calc // 
\calf \subseteq {\rm R}_m^{\cal F}(\calc)$) $\calf$ lacks the power to 
code the input into its output in a way that $\calc$ can decode.  However,
for flexible language classes such as P, NP, etc., and flexible
function classes such as \#P, OptP~\cite{kre:j:optimization}, etc.,
this equality will hold, as has been noted by Hemaspaandra
and Hoene~\cite{hem-hoe:j:nnt} for the particular case of
${\rm R}_m^{\rm OptP}(\parityp)$.  In particular, in 
terms of relations of
interest in the current paper, note that we clearly can claim
$$\psnplus = \np // \sharpp = \npclosure, \mbox{ and}$$
$$\psone = \p // \sharpp = \pclosure.$$
Note that expressions of this form are perhaps more
natural than one might first guess.  They capture the 
power of a class whose computation is aided by some 
advice provided by some complexity-bounded function
class.  Indeed, $\red{m}{\rm OptP}{\rm P}$ turns
out, as Krentel established~\cite{kre:j:optimization}, to 
be exactly $\pnp$.  $\red{m}{\rm OptP}{\parityp}$ turns out 
to be exactly the sets that $\leq_m^p$-reduce  to languages
having easy ``implicit membership tests''~\cite{hem-hoe:j:nnt}, and 
also is exactly the class of languages accepted at the 
second level of Cai and Furst's~\cite{cai-fur:j:bottleneck}
safe storage hierarchy
(\cite{ogi:j:serializable}, see also the 
discussion in~\cite{hem-ogi:j:ssf}).

Generalizing the earlier definition
to the case of more queries, 
for any function or language classes 
$\calone$ and $\caltwo$ and for any
$k$, 
let $\p^{{\cal C}_1[1]:{\cal C}_2[k]}$ 
denote the class of languages 
accepted by polynomial-time machines making 
at most one query to a $\calc_1$ oracle 
followed by at most $k$ 
queries to a $\calc_2$ oracle.  
$\p^{{\cal C}_1[1]:{\cal C}_2[{\cal O}(1)]}$ will denote 
$\bigcup_{k>0}
\p^{{\cal C}_1[1]:{\cal C}_2[k]}$.

We say that an NPTM is {\em normalized\/} if, for some polynomial $p$
and for every input $x$, every computation path on input $x$ has
exactly $p(|x|)$ nondeterministic choices.  (Note that it is not the
case that every computation step must make a nondeterministic choice.  However,
under our definition, on each input $x$ every possible binary string of
exactly $p(|x|)$ bits corresponds uniquely to a computation path.)
Though normalization is known not to be important
in defining NP languages or even---as shown by
Simon~\cite{sim:thesis:complexity}---PP languages, there is
evidence
that the type of
normalization one uses is critical in defining BPP 
languages~\cite{han-hem-thi:j:threshold}.  Here, it
will be important in our definition of SelfPath.  However, 
we note that in our
definition of SelfOutput, 
even if a
normalization requirement were added the class defined would 
remain the same.

We now introduce the classes SelfOutput and SelfPath, which 
model self-speci\-fy\-ing acceptance.  A language $L$ is in SelfOutput if for 
some NPTM $M$ and all $x$, $x\in L$ if and
only if $\numaccm(x) \in iout_M(x)$.
A language $L$ is in SelfPath if for some normalized NPTM $M$ and every 
$x$, $x\in L$ if and only if (a)~$\numaccm(x)>0$ and
(b)~$\numaccm(x)  - 1 \in iacc_M(x)$, that is, the lexicographically 
$(\numaccm(x))$th path of $M(x)$ is an accepting path.
In the definition of SelfPath, we view the leftmost path 
(i.e., the path on which all nondeterministic guesses are 0) of the 
machine as the lexicographically first path, 
and so on.\footnote{It
is not hard to see that one forms the same class of languages with the 
following alternate definition, which in some sense starts the path 
counting at 0:
A language $L$ is 
in SelfPath if for some normalized NPTM $M$ and every $x$,
(a)~$M(x)$ has at least one 
rejecting path, and (b)~$x\in L \iff $ the lexicographically 
$(1+\numaccm(x))$th path of $M(x)$ is an accepting path.}

\section{Self-Specifying Acceptance Types}

We characterize $\selfpath$ and $\selfoutput$ as the sets that 
$\leq_m^{\rm \#P}$-reduce to P and NP sets, respectively.
\begin{theorem}\label{t:xx}~
\begin{enumerate}
\item \label{parta} $\selfpath = \pclosure$.
\item \label{partb} $\selfoutput = \npclosure$.
\end{enumerate}
\end{theorem}

\noindent{\bf Proof}~~~
{\bf (\ref{parta})} \quad Recall from Section~\ref{section:prelim} that 
$\pclosure = \psone$.
Suppose $A\in \selfpath$, then by definition there is an NPTM $M$ such that

$x\in A  \iff $the lexicographically $(\numaccm(x))$th path of $M(x)$ is 
an accepting path.
 
So with one call to a $\sharpp$ oracle we can compute the value of 
$\numaccm(x)$ and check whether the $(\numaccm(x))$th path of $M(x)$ is 
an accepting path. 
Since $M(x)$ is a complete tree, we can do this latter check in 
deterministic polynomial time.

Now let $A\in \psone$ via a deterministic oracle machine $M$ and, 
without loss of generality, let the $\sharpp$ oracle be $\sharpsat$. 
Without loss of generality we may assume 
that on each input $x$ 
it holds that $M$ asks exactly one question to $\sharpsat$, getting the 
answer $f(x)$.
 Furthermore, let $M'$ be an NPTM witnessing $f \in \sharpp$. 
Let $M'$ be normalized, i.e., for some polynomial $p$, $M'(x)$ makes exactly 
$p(|x|)$ nondeterministic moves.

\begin{figure}[t]
  \begin{center}
    \leavevmode
    \unitlength2.64mm
    \begin{picture}(48,25)
      \renewcommand{\arraystretch}{0.5}

      \put(24,20){\line( 6,-1){12}}
      \put(24,20){\line(-6,-1){12}}
      \multiput(12,18)(24,0){2}{\line( 3,-1){ 6}}
      \multiput(12,18)(24,0){2}{\line(-3,-1){ 6}}
      \multiput( 6,16)(12,0){4}{\line(-1,-2){ 5}}
      \multiput( 6,16)(12,0){4}{\line( 1,-2){ 5}}
      \multiput( 1, 6)(12,0){4}{\line( 1, 0){10}}

      \multiput( 6,16)(12,0){2}{\line( 0,-1){ 8}}
      \multiput( 6, 8)(12,0){2}{\line( 1,-2){ 1}}
      \put(30,16){\line(0,-1){10}}

      \put( 4,5.5){\makebox(0,0)[t]{$
          \begin{array}{c}
            \underbrace{\rule{6\unitlength}{0\unitlength}} \\
            \mbox{\footnotesize Paths of} \\
            \mbox{\footnotesize Case 1.1}
          \end{array}$}}
      \put( 9,5.5){\makebox(0,0)[t]{$
          \begin{array}{c}
            \underbrace{\rule{4\unitlength}{0\unitlength}} \\
            \mbox{\footnotesize Reject}
          \end{array}$}}

      \put(16,5.5){\makebox(0,0)[t]{$
          \begin{array}{c}
            \underbrace{\rule{6\unitlength}{0\unitlength}} \\
            \mbox{\footnotesize Paths of} \\
            \mbox{\footnotesize Case 2.1}
          \end{array}$}}
      \put(21,5.5){\makebox(0,0)[t]{$
          \begin{array}{c}
            \underbrace{\rule{4\unitlength}{0\unitlength}} \\
            \mbox{\footnotesize Accept}
          \end{array}$}}

      \put(27.5,5.5){\makebox(0,0)[t]{$
          \begin{array}{c}
            \underbrace{\rule{5\unitlength}{0\unitlength}} \\
            \mbox{\footnotesize Paths of} \\
            \mbox{\footnotesize Case 3.1}
          \end{array}$}}
      \put(32.5,5.5){\makebox(0,0)[t]{$
          \begin{array}{c}
            \underbrace{\rule{5\unitlength}{0\unitlength}} \\
            \mbox{\footnotesize Reject}
          \end{array}$}}

      \put(42,5.5){\makebox(0,0)[t]{$
          \begin{array}{c}
            \underbrace{\rule{10\unitlength}{0\unitlength}} \\
            \mbox{\footnotesize Reject}
          \end{array}$}}
    \end{picture}
  \end{center}
  \caption{Computation tree of $N(x)$.\label{fig:ctn}}
\end{figure}
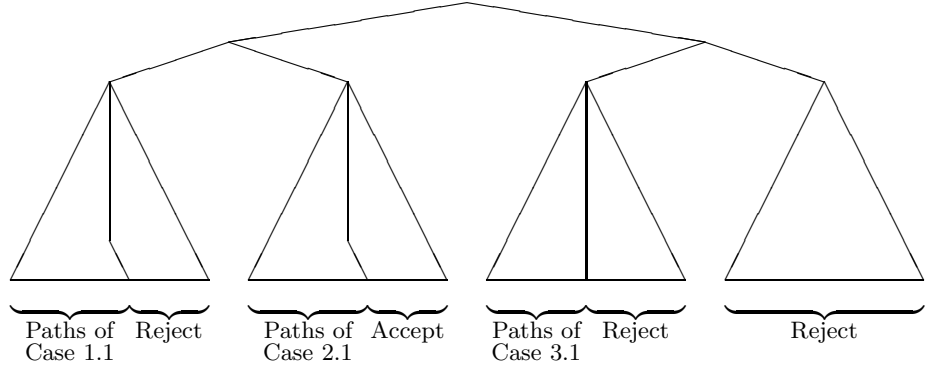

We will define an NPTM $N$, which will witness the fact that $A \in 
\selfpath$. 
$N(x)$ will be normalized to make exactly $p(|x|)+3$ nondeterministic 
moves.
The computation of $N(x)$ along path $y$, where $y \in 
\Sigma^{p(|x|)+3}$, 
goes as follows (see Figure~\ref{fig:ctn}). 
(Here we take strings $y$ to represent a computation path, 
where $y$ is simply the sequence of nondeterministic choices along the path.)
Recall that $0 \le f(x) \le 2^{p(|x|)}$, so $f(x)$ can take on $2^{p(|x|)}+1$ 
possible values.  The cases will reflect this (see the 
conditions defining Cases~1.1 and~2.1, which lets $int(y')$ sweep through
exactly those values).

\begin{description}
\item[Case 1] $y= 00y'$.
  \begin{description}
    \item[Case 1.1] $int(y') \le 2^{p(|x|)}$. \\Simulate $M(x)$, assuming the 
answer to the 
$\sharpsat$ query is $int(y')$ 
and reject 
if and only if $M(x)$ accepts.
    \item[Case 1.2] $int(y') > 2^{p(|x|)}$. \\Reject.
    \end{description}
\item[Case 2] $y=01y'$.
  \begin{description}
    \item[Case 2.1] $int(y') \le 2^{p(|x|)}$. \\Simulate $M(x)$, assuming the 
answer to the 
$\sharpsat$ query is $int(y')$ 
and accept if and 
only if $M(x)$ accepts.
    \item[Case 2.2]  $int(y') > 2^{p(|x|)}$. \\Accept.
  \end{description}
\item[Case 3] $y=10y'$.
  \begin{description}
    \item[Case 3.1] $y'=0z$, where $z \in \Sigma^{p(|x|)}$.\\ 
Simulate $M'(x)$ on path $z$, i.e., accept if and only if path $z$ on 
$M'(x)$ accepts.
    \item[Case 3.2] $y'=1z$, where $z \in \Sigma^{p(|x|)}$. \\Reject.
  \end{description}
\item[Case 4] $y=11y'$.\\
Reject. 
\end{description}

\noindent

\begin{tabbing} Note that \qquad\= \makebox{$||\{0y'\condition y'\in 
\Sigma^{p(|x|)+2}$ and $N(x)$ accepts along $0y'\}||=
2^{p(|x|)+1}$}\\
and \>\makebox{$||\{1y'\condition y'\in \Sigma^{p(|x|)+2}$ and 
$N(x)$ accepts along $1y'\}||=f(x)$}.
\end{tabbing}

Hence $N(x)$ has exactly \mbox{$2^{p(|x|)+1}+f(x)$} {\em accepting\/} paths. 
However, note that the \mbox{$(2^{p(|x|)+1}+f(x))$th {\em path\/}} of $N(x)$ 
is a path of the form of Case~2.1 and by construction
it accepts if and only if $M(x)$, 
given the answer $f(x)$, accepts. So $A \in \selfpath$.

\medskip

{\bf (\ref{partb})} \quad Recall from Section~\ref{section:prelim} 
that $\npclosure = \psnplus$. Suppose $A \in \selfoutput$ then by definition 
there is an NPTM $M$ such that

$ x \in A \iff \numaccm(x) $ is the integer
reading of the output of some accepting path of $M(x)$.

So with one call to $\sharpsat$ we can compute the value of $\numaccm(x)$ 
and by querying \mbox{``$\langle x,string(\numaccm(x)) \rangle \in 
B $?''}, where 
$B=_{def}\{\langle x, string(m) \rangle \condition m \in iout_M(x)\}$, we
can decide whether $x \in A$. Note that $B \in \NP $ and that the second 
query is a positive one.

Now let $A \in \psnplus$ and thus $A \in {\rm P}^{{\sharpsat}[1]:{\sat}[1]_+}$.
Furthermore we may, without loss of 
generality (see Footnote~\ref{f:lg}), assume that some deterministic 
machine, $M$, that witnesses
\mbox{$A \in {\rm P}^{{\sharpsat}[1]:{\sat}[1]_+}$} has the 
property that on every input it asks exactly one query to each oracle and  
the query to $\sat$ 
is strictly positive in the sense that the machine accepts if and only if 
the query is answered ``yes.''
This strict positivity property is important to 
keep
in mind throughout the proof.

Since $\sharpp$ is closed under polynomial-time input transformation let 
$f(x)$ be the answer to the query that $M(x)$ makes to $\sharpsat$, and denote 
the question to $\sat$ by $z(x,f(x))$. Note that, 
as $\sat$ is a cylinder, we may
without loss of generality assume that $z$ is such that there 
exists a polynomial $q$ for which
$(\forall x \in \sigmastar)
(\forall n : 0\leq n \leq 2^{p|x|)})[|z(x,n)|=q(|x|)]$.

Let $M'$ be an NPTM witnessing $f \in \sharpp$. Let $M'$ be normalized to 
have branching depth exactly given by the polynomial $p$. Suppose $M_{\sat}$ 
witnesses $\sat \in \NP$ and $M_{\sat}$ is normalized to have branching depth 
exactly given by the polynomial $q'$, so $M_{\sat}(z(x,f(x)))$ has exactly 
$2^{q'(q(|x|))}$ paths. Without loss of generality, let $q$ and $q'$ be such 
that\\
\mbox{~~~~~~~~~~~~~~~~~$(\forall n) [q(n) \ge 1$ and $q'(n) \ge 1]$.}

\begin{figure}[t]
\begin{center}
  \leavevmode
  \unitlength2.648mm
  \begin{picture}(48,32)
     \put(24,30){\line( 6,-1){12}}
     \put(24,30){\line(-6,-1){12}}
     \multiput(12,28)(24,0){2}{\line(-1,-1){10}}
     \multiput(12,28)(24,0){2}{\line( 1,-1){10}}
     \multiput( 2,18)(24,0){2}{\line( 1, 0){20}}
     \multiput(36,28)(0,-4){3}{\line(-1,-2){ 1}}
     \multiput(35,26)(0,-4){2}{\line( 1,-2){ 1}} 
     \multiput(35,18)(0,-3){2}{\line(-1,-2){ .75}}
     \multiput(34.25,16.5)(0,-3){2}{\line( 1,-2){ .75}}
     \put(35,12){\line( 1,-2){ .75}}
     \put(35.75,10.5){\line(-1,-2){ 3}}
     \put(32.75, 4.5){\line( 1,-2){ .75}}
     \put(35,12){\line(-1,-1){ 9}}
     \put(35,12){\line( 1,-1){ 9}}
     \put(26, 3){\line( 1, 0){18}}
     \put(33.5, 3){\line( 0,-2){2}}

     \put(24,31.5){\makebox(0,0)[t]{Input $x$}}
      \put(12,22){\makebox(0,0)[t]{$M'(x)$}} 
      \newlength{\breite}
      \setlength{\breite}{10\unitlength}
      \put(12,17.5){\makebox(0,0)[t]{$
          \begin{array}{c}
            \underbrace{\rule{2\breite}{0\unitlength}} \\
            \mbox{$f(x)$ accepting paths each outputting 0}
          \end{array}$}}

      \put(39,20){\makebox(0,0)[t]{guess $n$}}
      \put(35,16.5){\makebox(0,0)[tl]{
            \parbox{11\unitlength}{\baselineskip0.3cm 
                                   $M(x)$, assuming first answer was $n$}}}
      \put(37,12){\makebox(0,0)[tl]{$z(x,n) \in \sat$?}}
      \put(33.45,5.5){\makebox(0,0)[tl]{$M_{\rm SAT}(z(x,n))$}}
      \put(35, 1){\makebox(0,0)[t]{accept and output appropriate value}} 
  \end{picture}
\end{center}
\caption{Computation tree of $T(x)$.\label{fig:ctt-tree-2}}
\end{figure}
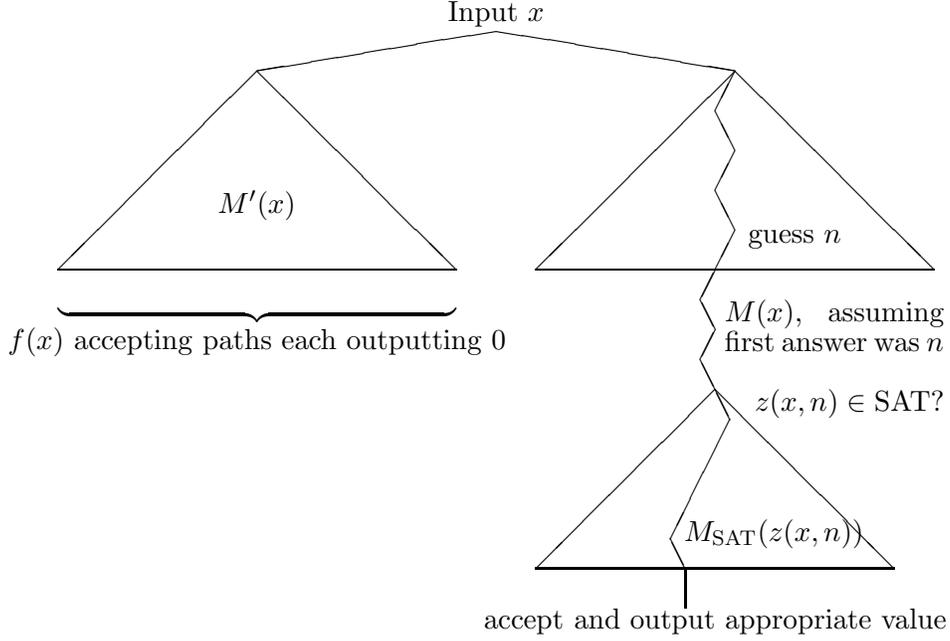

We construct the NPTM $T$ witnessing $A \in \selfoutput$. On input $x$, 
$T(x)$ does the following (see Figure~\ref{fig:ctt-tree-2}):

\begin{itemize}
\item $T(x)$ guesses $i \in \{0,1\}$.
\begin{itemize}
\item If $i=0$ was guessed, $T(x)$ runs $M'(x)$, but outputs 
the integer~0 on all accepting 
paths of $M'(x)$.

\item If $i=1$ was guessed, $T(x)$ guesses an $n$ , $0 \le n \le 2^{p(|x|)}$, 
and simulates what $M(x)$ would do were $n$ the answer to the 
$\sharpsat$ query.
$M(x)$ under this assumption queries ``$z(x,n) \in \sat$?''~so $T(x)$ 
guesses a path $y$ of $M_{\sat}(z(x,n))$. 
Recall that $M$'s behavior is such that it accepts 
if and only if the answer to its second query is ``yes.''
If $M_{\rm SAT}(z(x,n))$ accepts on path $y$, 
then $T(x)$ also accepts on path $y$ and outputs 
the integer
\mbox{ $(2^{p(|x|)}+1)2^{q'(q(|x|))}+n$}. Otherwise $T(x)$ accepts on 
the present path and outputs the integer~$0$.
\end{itemize}
\end{itemize}
By construction $T(x)$ has exactly $(2^{p(|x|)}+1)2^{q'(q(|x|))}+f(x)$ 
accepting paths. This value is an output of some accepting path, if and 
only if $x \in A$. So $A \in \selfoutput$.~\qed

Note that the proof technique in fact can be applied to characterize via 
self-specifying acceptance other 
classes, such as 
$\rms(\up)$,
$\rms(\parityp)$,
$\rms(\ceqp)$, and
$\rms(\pp)$.
For example, if we change the definition of SelfOutput to add a requirement 
that no two
accepting paths output the same value, this ``unambiguous SelfOutput'' class 
equals $\rms(\up)$.
Similarly, if we change the definition of SelfOutput to 
accept exactly when the number of paths itself is the output on 
an odd number of paths, we obtain $\rms(\parityp)$.
Similar claims hold for $\rms(\ceqp)$ and (with a bit of care)
for $\rms(\pp)$.

Let us adopt Valiant's~\cite{val:j:permanent} standard definition of
$\sharpnp$ (informally, $\sharpnp = (\sharpp)^{\np}$), and the
analogous definition of $\sharpph$ (informally, $\sharpph =
(\sharpp)^{\ph}$)~\cite{tod-wat:j:one-call}.\footnote{Vollmer~\protect\cite{vol:thesis:functions}
and Toda 
and Watanabe 
(\cite{tod-wat:j:one-call},
using the different notation ``$\mbox{NUM} \cdot {\cal C}$'')
have proposed interesting and different ``$\#$''-type classes.
Often, though not always, Vollmer's classes are referred to 
using notation such as $\# \cdot \np$, in order to avoid 
ambiguity as to whether his classes or Valiant's classes
are being discussed (see~\cite{hem-vol:j:satanic}).  
Here, we uniformly use Valiant's classes, though we mention
in passing that for $\sharpph$ the two notions 
are known to coincide~\protect\cite[Proposition~3.1]{tod-wat:j:one-call}.}
In this paper, we have discussed the sets 
$\manyonesharp$-reducible to certain classes.
One might naturally wonder 
whether $\leq_m^{\sharpnp}$-reductions to the same classes
yield even greater computational power.
However, note that from Toda and Watanabe's~\cite{tod-wat:j:one-call} result 
$\sharpph \subseteq {\rm F}\psone$ we can easily prove
the 
following proposition, which says that 
for most natural classes
$\leq_m^{\sharpnp}$-reductions to the class
(or even $\leq_m^{\sharpph}$-reductions to the class)
yield no greater computational power than 
$\manyonesharp$-reductions to the class.
(We say {\em $\cal C$ is closed downwards under 
$\manyone$ reductions} if ${\rm R}_m^p({\cal C})
\subseteq {\cal C}$.)

\begin{proposition}\label{p:tw}
For any complexity class $\calc$ closed downwards
under $\manyone$ reductions,
it holds that 
$\redmsharp{\calc}={\rm R}_m^{\sharpph}(\calc)$.
\end{proposition}
  
\smallskip

\noindent
{\bf Proof: }
     The inclusion 
     $\redmsharp{\calc} \subseteq {\rm R}_m^{\sharpph}(\calc)$ is immediate.
     For the ``$\supseteq$'' 
        inclusion, let $A \in {\rm R}_m^{\sharpph}(\calc)$ via 
     $f \in \sharpph$ and $C \in \calc$. Thus,
     $x \in A \iff string(f(x)) \in C$.   Note that due to the  
     inclusion $\sharpph \subseteq {\rm F}\psone$~\cite{tod-wat:j:one-call}, 
     we 
     also have $string(f(\cdot)) 
     \in {\rm F}\psone$. Hence there exist a DPTM $M$ and a function 
     $g \in \sharpp$ such that, for each $x \in \sigmastar$,
     $M^{g[1]}(x)$ computes $string(f(x))$.  
Without loss of generality we assume that $M(x)$ always asks exactly one
question to its oracle.

Let $q(x)$ denote the query asked by $M(x)$.
For any string $x \in \sigmastar$, $(1x)_{\rm binary}$ 
will denote the integer value of the string formed by 
prepending a 1 before $x$ and interpreting the string as a binary
integer.
Let $\langle
\cdot , \cdot \rangle_{\naturalnumbers}$ be an 
easily computable pairing function from
$\naturalnumbers \times \naturalnumbers$ to $\naturalnumbers$ 
such that 
(a)~$h(x) = \langle (1x)_{\rm binary} , g(q(x)) \rangle_{\naturalnumbers}$
is a $\sharpp$ function, and 
(b)~there exist polynomial-time functions to extract from $\langle \cdot,
\cdot \rangle_{\naturalnumbers}$ its 
first and second components.
Such pairing functions are known to exist via
standard techniques~\cite{pap-zac:c:two-remarks,cai-hem:j:parity}.
For any $m \in \naturalnumbers$, let $\widehat{m}$ denote 
the constant function that, on any input, returns the value $m$.
Let 
$$D=\{ string( \langle (1x)_{\rm binary}, m\rangle_{\naturalnumbers} )
\condition 
\mbox{ the value output by }M^{\widehat{m}}(x) 
     \mbox{ is in }C\},$$ and note that 
     $D \in \calc$ since $\calc$ is closed downwards
under $\manyone$ reductions.
We have 
     \begin{eqnarray*}
     x \in A &\iff& string( f(x) ) \in C\\
             &\iff& string( h(x) ) \in D.
     \end{eqnarray*}
     
     This shows $A \in \redmsharp{\calc}$.~\qed

Proposition~\ref{p:tw}  implies that 
if in the definition of $\selfoutput$ we 
replace 
``$x \in L \iff \#acc_M(x)$ is the integer 
reading of the output of some accepting path'' 
with 
``$x \in L \iff span_M(x)$ is the integer 
reading of the output of some accepting path''
(where $span_M(x)$~\cite{koe-sch-tor:j:spanp} as is 
standard denotes the 
number of distinct outputs of $M(x)$),
then the class defined remains unchanged.
 
From Theorem~\ref{t:xx} and the 
discussion of Section~\ref{section:prelim}, it is clear that we have 
the following.
\begin{corollary}\label{c:oc}
$\selfpath = \selfoutput \iff \psone = \psnplus$.
\end{corollary}

Are SelfPath and SelfOutput in fact equal?  
In light of our Corollary~\ref{c:oc}, this is equivalently 
a question---$\psone = \psnplus$---about somewhat
more familiar-looking complexity classes, 
and in this form it allows us to see more clearly 
a relationship with an open issue from the literature.
In particular,
no oracle
separation yet exists for these classes.  In fact, 
separating even over a vastly larger gap 
that includes these classes
is an open issue.  That is,
$\pp^{\parityp} \subseteq \psone \subseteq \psnplus \subseteq \p^{{\rm
\#P}[2]} \subseteq \pspace$ (the first containment---which 
is nontrivial---is due to 
Toda~\cite{tod:thesis:counting}), yet no known oracle separates
$\pp^{\parityp}$ from PSPACE (see Green~\cite{gre:c:equals-mods} for
background and for progress on a related line).  
On the other hand, 
as even 
$\p^{ {\rm NP} [{\cal O}(\log n)]: {\rm \#P}[1]}$ easily equals
$\psone$ 
(via the technique of~\cite{cai-hem:j:parity,pap-zac:c:two-remarks}), 
clearly if
query order can be swapped in the characterization of SelfOutput then
$\selfpath = \selfoutput$.  

It would be nice to
achieve some structural collapse from the assumption $\psone =
\psnplus$.
We in fact can prove such a collapse.
\begin{theorem}\label{t:down}
$\selfpath = \selfoutput \iff     \psone=\psnnoplusbigohone$.
\end{theorem}

Theorem~\ref{t:down} equivalently says
$$\psone = \psnplus 
\iff     \psone=
\psnnoplusbigohone.$$
Related work shows that 
the equality of $\selfpath$ and $\selfoutput$ 
would collapse the boolean hierarchy over 
$\selfoutput$~\cite{hem-wec:t:sharpp-reductions}.

\noindent 
{\bf Proof of 
Theorem~\protect\ref{t:down}:  }
We will prove this in two steps.  
First we will prove that 
$$\psone = \psnplus \iff
\psone = 
\p^{\sharpp[1]:\np[1]},$$
and then we will prove that 
$$\psone = 
\p^{\sharpp[1]:\np[1]}
\iff
\psone = \psnnoplusbigohone.$$
From these two equivalences we are done, 
in light of Corollary~\ref{c:oc}.

The right to left implication of 
$$\psone = \psnplus \iff
\psone = 
\p^{\sharpp[1]:\np[1]}$$ is immediate.  
Regarding the left to right implication, 
suppose that $B \in 
\p^{\sharpp[1]:\np[1]}$ and assume that 
$\psone = \psnplus$.  
Note that 
$$\p^{\sharpp[1]:\np[1]} \subseteq 
\{ L_1 \cup L_2 \condition L_1 \in \psnplus \land 
\overline{L_2} \in \psnplus\}.$$
From this, combined with 
our $\psone = \psnplus$ assumption and the fact that 
$\psone$ is closed under complementation and union,
we have $B \in \psone$.  

That 
$\psone = 
\p^{\sharpp[1]:\np[1]}
\iff
\psone = \psnnoplusbigohone$
holds can be seen as 
follows.  
The right to left implication is immediate.  
Regarding the left to right implication
we have, as our assumption, that 
$\psone = 
\p^{\sharpp[1]:\np[1]}$.  We will prove 
that, for each $k \geq 1$:
$\psone = 
\p^{\sharpp[1]:\np[k]} \implies
\psone = 
\p^{\sharpp[1]:\np[k+1]}$.  This suffices, via induction.

So let $k$ be some positive integer and assume that 
$\psone = 
\p^{\sharpp[1]:\np[k]}$.  Let $A \in 
\p^{\sharpp[1]:\np[k+1]}$.  Let $M$ be a deterministic
polynomial-time machine 
that accepts $A$ via at most one query to $\sharpp$ followed by 
at most $k+1$ 
queries to $\np$.  Without loss of generality, assume that 
$M$ always asks exactly one query to $\sharpsat$ followed by 
exactly $k+1$ queries to $\sat$.  
Consider the machine $M_{no}$ that (on each input) exactly simulates
$M$, except rather than asking the first of $M$'s $k+1$ queries 
to $\sat$ it assumes the answer to that query is ``no.''
Consider also the machine $M_{yes}$ that (on each input) exactly simulates
$M$, except rather than asking the first of $M$'s $k+1$ queries 
to $\sat$ it assumes the answer to that query is ``yes.''
Let $$L_{no} = \{x \condition x \in L(M_{no}^{\sharpsat[1]:\sat[k]}) \}$$
and
$$L_{yes} = \{x \condition x \in L(M_{yes}^{\sharpsat[1]:\sat[k]}) \}.$$
Note that $L_{no} \in 
\p^{\sharpp[1]:\np[k]}$ and 
$L_{yes} \in \p^{\sharpp[1]:\np[k]}$.
By our assumption, it follows that both $L_{no}$ and $L_{yes}$
are in $\psone$.  Let $\widehat{M}_{no}$ 
(respectively, $\widehat{M}_{yes}$) be a machine that makes one
query to $\sharpsat$, and whose language is
$L_{no}$ (respectively, $L_{yes}$).
We claim that $A \in 
\p^{\sharpp[1]:\np[1]}$, via the following 
machine $M_v$.  On input $x$, $M_v(x)$ asks its $\sharpp$ oracle 
for the answer to (a)~the $\sharpsat$ query asked by $M(x)$,
(b)~the $\sharpsat$ query asked by $\widehat{M}_{no}(x)$, and 
(c)~the $\sharpsat$ query asked by $\widehat{M}_{yes}(x)$.
This can all be done via {\em one\/} query to $\sharpsat$, as 
it is well known that three parallel 
queries to $\sharpsat$ can be
encoded as one query to $\sharpsat$~(\cite{pap-zac:c:two-remarks},
see also~\cite{cai-hem:j:parity}).  Using the answer to 
item~(a), $M_v(x)$ then determines the first query to $\sat$
that would be made by $M(x)$, namely, the $\sat$ query $M(x)$ makes 
after $M(x)$ gets the reply from its $\sharpsat$ query.
$M_v(x)$ asks this question to its own $\sat$ oracle.  
If the answer is ``no,'' then $M_v(x)$ accepts if and only if 
$x \in L_{no}$, and we can easily evaluate this with no
additional queries, via simulating $\widehat{M}_{no}(x)$ using
the $\sharpsat$ answer obtained in item~(b).
If the answer is yes, then $M_v(x)$ accepts if and only if 
$x \in L_{yes}$, and we can easily evaluate this with no
additional queries, via simulating $\widehat{M}_{yes}(x)$ using 
the $\sharpsat$ answer obtained in item~(c).
So $A \in 
\p^{\sharpp[1]:\np[1]}$, and thus by our 
assumption we have
$A \in 
\p^{\sharpp[1]}$.~\qed

\section{Open Questions}

We completely characterized 
the self-specifying
classes SelfOutput and SelfPath as the
sets $\leq_m^{\rm \#P}$-reducible
to
NP and P sets, respectively.  Can one prove $\selfoutput = \selfpath$?  That
is, can one prove $\pclosure = \npclosure$, which would instantly imply 
by Theorem~\ref{t:down}  
that 
$\p^{ {\rm \#P}[1]} =\psnnoplusbigohone$
(equivalently, 
$\p^{{\rm NP} [{\cal O}(1)]: {\rm \#P}[1]} =\psnnoplusbigohone$)?

Though Simon proved that the class PP remains the same regardless of
whether or not the underlying machines are 
normalized,\footnote{Note that for 
probabilistic 
polynomial-time 
machines, normalization in terms of 
{\em number of choices per path in the model in which not every step 
is necessarily viewed as a choice node\/} and
normalization in terms of 
{\em number of choices per path in the model in which every step 
is necessarily viewed as a choice node\/} are clearly equivalent.
This holds true in the unbounded-error case.  It is also true
in the bounded-error case (i.e., BPP is the same class
whether one defines it using the former normalization or the latter 
normalization).~~In particular, for PP 
it does not matter which normalization model one adopts;  for PP,
both are equivalent to the unnormalized model.}
a number of
recent papers have studied normalization in other
contexts
and it is now clear that in some contexts---e.g., the class
BPP---classes may be different depending on whether or not there is a
normalization 
requirement~\cite{han-hem-thi:j:threshold,jen-mck-the:j:leaf,her-vol-wag:j:unbalanced-trees}.  Recall that SelfOutput does remain the 
same whether defined with 
or without the requirement that the underlying machines be normalized.
Does SelfPath remain the same class if its normalization requirement 
is removed?  Clearly the resulting class contains SelfPath, i.e., 
$\psone$, and is contained in $\psharpp$ (which Vollmer 
and Wagner~\cite{vol-wag:j:middle}
showed 
equals $\p^{\mbox{\scriptsize{}``NameofMiddlePath''}[1]}$;
thus
$\p^{\mbox{\scriptsize{}``NameofMiddlePath''}[1]}$, though it 
is an upper bound for the unnormalized version of 
$\selfpath$, does not offer a tighter 
upper bound than $\psharpp$).

{\samepage
\begin{center}
{\bf Acknowledgments}
\end{center}
\nopagebreak
\indent
We are very grateful to Edith Hemaspaandra, Johannes
K\"obler, J\"org Rothe, and an anonymous
referee for 
helpful suggestions.
}

\bibliographystyle{alpha}

\end{document}